\documentclass[lettersize,journal]{IEEEtran}
\usepackage{amsmath,amsfonts}
\usepackage{algorithmic}
\usepackage{algorithm}
\usepackage{array}
\usepackage[caption=false,font=normalsize,labelfont=sf,textfont=sf]{subfig}
\usepackage{textcomp}
\usepackage{stfloats}
\usepackage{amssymb}
\usepackage{cleveref} 
\usepackage{url}
\usepackage{cite}
\usepackage{epsfig}
\usepackage{epstopdf}
\usepackage{verbatim}
\usepackage{caption}
\usepackage{graphicx}
\def\BibTeX{{\rm B\kern-.05em{\sc i\kern-.025em b}\kern-.08em
		T\kern-.1667em\lower.7ex\hbox{E}\kern-.125emX}}
\usepackage{balance}

\begin{document}
	\title{Passive Six-Dimensional Movable Antenna (6DMA)-Assisted Multiuser Communication}
	\author{Haozhe Wang, Xiaodan Shao,~\IEEEmembership{Member,~IEEE,} Beixiong Zheng,~\IEEEmembership{Senior Member,~IEEE,} Xiaoming Shi, and Rui Zhang,~\IEEEmembership{Fellow,~IEEE} \vspace{-25pt}
	\thanks{H. Wang is with School of
	Science and Engineering, The Chinese University of Hong Kong, Shenzhen,
	Guangdong, 518172, China (e-mail: haozhewang1@link.cuhk.edu.cn).}
	\thanks{Xiaodan Shao is with Department of Electrical and Computer Engineering, University of Waterloo, Waterloo, ON N2L 3G1, Canada (e-mail: x6shao@uwaterloo.ca). \emph{(Corresponding author: Xiaodan Shao)}}
	\thanks{B. Zheng is with School of Microelectronics, South China University of
		Technology, Guangzhou 511442, China (e-mail: bxzheng@scut.edu.cn).}
	\thanks{X. Shi is with School of
		Science and Engineering, The Chinese University of Hong Kong, Shenzhen,
		Guangdong 518172, China (e-mail: xiaomingshi@link.cuhk.edu.cn).}
    \thanks{R. Zhang is with School of Science and Engineering, Shenzhen Research
    	Institute of Big Data, The Chinese University of Hong Kong, Shenzhen,
    	Guangdong 518172, China. He is also with the Department of Electrical and
    	Computer Engineering, National University of Singapore, Singapore 117583
    	(e-mail: elezhang@nus.edu.sg).}
}
		
	\maketitle
	
	\begin{abstract}
		Six-dimensional movable antenna (6DMA) is a promising solution for enhancing wireless network capacity through the adjustment of both three-dimensional (3D) positions and 3D rotations of distributed antenna surfaces. Previous works mainly consider 6DMA surfaces composed of active antenna elements, thus termed as active 6DMA. In this letter, we propose a new passive 6DMA system consisting of distributed passive intelligent reflecting surfaces (IRSs) that can be adjusted in terms of 3D position and 3D rotation. Specifically, we study a passive 6DMA-aided multiuser uplink system and aim to maximize the users' achievable sum rate by jointly optimizing the 3D positions, 3D rotations, and reflection coefficients of all passive 6DMA surfaces, as well as the receive beamforming matrix at the base station (BS). To solve this challenging non-convex optimization problem, we propose an alternating optimization (AO) algorithm that decomposes it into three subproblems and solves them alternately in an iterative manner. Numerical results are presented to investigate the performance of the proposed passive 6DMA system under different configurations and demonstrate its superior performance over the traditional fixed-IRS counterpart for both directive and isotropic radiation patterns of passive reflecting elements.
	\end{abstract}

	\section{Introduction}
	\IEEEPARstart{I}{ntelligent} reflecting surface (IRS) has emerged as a promising technology for creating smart and reconfigurable wireless communication environments in a cost-effective manner \cite{wu2024intelligent}. Specifically, IRS is a planar surface composed of massive reflecting elements (REs), each of which can individually adjust its own reflection coefficient, thereby collaboratively altering the strength/direction of the reflected signal for achieving various functions, such as signal enhancement, wireless sensing \cite{shao2022target, shaomoun}, and interference nulling \cite{hua}. However, most of the existing works focus solely on the design of reflection coefficients, ignoring the spatial degrees of freedom (DoFs) associated with the position and rotation of IRS. Thus, by fully exploiting these new spatial DoFs, further performance enhancement over traditional IRS with fixed position/rotation is anticipated. 

Recently, six-dimensional movable antenna (6DMA) system composed of multiple position/rotation adjustable antenna surfaces has been proposed as a new and effective solution for improving the wireless network capacity \cite{shao20246d, discrete}. By jointly optimizing the three-dimensional (3D) positions and rotations of all 6DMA antenna surfaces in a given site space, the 6DMA-equipped  transmitters/receivers can adaptively allocate antenna resources to maximize the array gain and spatial multiplexing gain according to the users' distribution. In practice, the positions and rotations of 6DMA surfaces can be adjusted either continuously\cite{shao20246d} or discretely \cite{discrete, shao20246dma}, and they can be optimized with or without prior knowledge of the spatial channel distribution of the users in the network \cite{shao20246d, discrete}. In addition to communication, the authors in \cite{censing} proposed a 6DMA-aided wireless sensing system. Besides, the authors of \cite{jstsp} proposed instantaneous and statistical channel estimation algorithms tailored for 6DMA systems by exploiting a new directional sparsity of 6DMA channels. In contrast to the existing fluid antenna system, or movable antenna system \cite{new2023fluid, zhu}, which only considers antenna's movement on a given 2D surface or a given line with their fixed/given rotations to mitigate the impact of channel deep fading, 6DMA can fully exploit the spatial variation of wireless channels via joint 3D position/rotation optimization as well as flexibly allocating antenna resources to match the users' channel distribution \cite{free6DMA}. 
	
However, existing 6DMA studies primarily focus on enhancing the direct link between the active 6DMA-equipped base station (BS) and its served users \cite{ming}, whose performance can be dramatically degraded due to the potential blockage. In view of this, IRS-aided wireless communication systems can exploit IRS reflection links to bypass obstacles between the BS and users. Motivated by this as well as the significant performance gains achieved by 6DMA's new spatial DoFs, we propose in this letter a new passive 6DMA-aided system consisting of distributed passive IRSs rather than active antennas/arrays, to improve the wireless network capacity and reduce communication outage. By enabling IRSs with the full control of 3D position and 3D rotation, passive 6DMA (also called 6D-IRS) allows for more flexible and adaptive reconfiguration of wireless channels between the BS and users, thus enhancing their communication performance, particularly in the presence of blockages that impair the direct links between them.  
	
		More specifically, we consider a passive 6DMA-assisted multiuser uplink communication system with the goal of maximizing the users' achievable sum rate by jointly optimizing the positions, rotations, and reflection coefficients of all passive 6DMA surfaces as well as the receive beamforming matrix at the BS. We first model the radiation pattern of REs in 6D-IRS channel and investigate its effect on system rate performance. Then, to tackle the non-convex optimization problem, an alternating optimization (AO) algorithm is proposed to decompose it into three subproblems and solve them alternately in an iterative manner. Simulation results are presented to evaluate the performance of passive 6DMA in different configurations and demonstrate its superiority over the traditional fixed-IRS scheme in terms of achievable sum rate. 
 	\section{System Model and Problem Formulation}
	\subsection{Channel Model}
\begin{figure}[!t]
	\centering
	\includegraphics[width=0.4\textwidth]{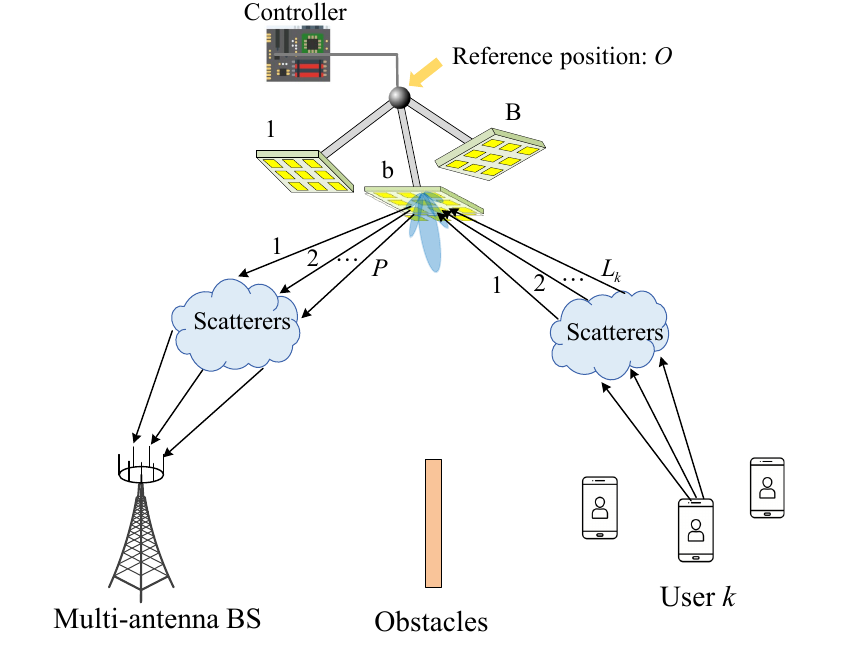}
	\caption{Passive 6DMA-assisted multiuser uplink communication.}
	\label{Fig_main}
\end{figure}

As shown in Fig. \ref{Fig_main}, we consider the uplink transmission of a multiuser communication system, where $B$ passive 6DMA surfaces are deployed to enhance the communications from $K$ single-antenna users to a BS equipped with $M$ antennas. Each passive 6DMA surface is assumed to be a uniform planar array (UPA) consisting of $N=N_xN_y$ REs, with $N_x$ and $N_y$ denoting the numbers of REs in a row and column, respectively. The passive 6DMA surfaces are connected to the central processing unit (CPU) via extendable and rotatable rods embedded with flexible wires, and thus their 3D positions/rotations and reflection coefficients can be adjusted cooperatively by the CPU. We assume that the direct link between each user and the BS is severely blocked by obstacles and thus is ignored. 

Let $\boldsymbol{g}_k\left(\boldsymbol{q}_b, \boldsymbol{u}_b\right)\in\mathbb{C}^{N\times1}$ and $\boldsymbol{V}\left(\boldsymbol{q}_b, \boldsymbol{u}_b\right)\in\mathbb{C}^{M\times N}$ denote the baseband equivalent channels for user $k$$\rightarrow$passive 6DMA surface $b$ and passive 6DMA surface $b$$\rightarrow$BS links with respect to (w.r.t.) surface $b$'s position $\boldsymbol{q}_b$ and rotation $\boldsymbol{u}_b$, respectively. Let $\boldsymbol{q}_b\triangleq \left[x_b, y_b, z_b\right]^T\in \mathbb{R}^{3\times1}$ and $\boldsymbol{u}_b\triangleq\left[\zeta_{x,b}, \zeta_{y,b}, \zeta_{z,b}\right]^T \in\mathbb{R}^{3\times1}$ denote the position and rotation of passive 6DMA surface $b$, with $b\in\mathcal{B}\triangleq\left\{1,\cdots,B\right\}$, where $x_b$, $y_b$, and $z_b$ denote the coordinates of the center point of the passive 6DMA surface $b$ in the global Cartesian coordinate system (CCS) $O$-$xyz$, with the reference position shown in Fig. \ref{Fig_main} serving as the origin $O$; $\zeta_{x,b}\in\left[0,2\pi\right)$, $\zeta_{y,b}\in\left[0,2\pi\right)$, and $\zeta_{z,b}\in\left[0,2\pi\right)$ represent the rotation angles w.r.t. the $x$-axis, $y$-axis, and $z$-axis, respectively. Let $\boldsymbol{\theta}_b\triangleq\left[\theta_{b,1},\cdots\theta_{b,N}\right]^T$ denote the equivalent reflection coefficients of passive 6DMA surface $b$, where the reflection amplitude is set to one to maximize the signal reflection power and thus $\left|\theta_{b,n}\right|=1$, $\forall n=1\cdots N$, $b\in \mathcal{B}$. With the above setup, the cascaded channel from user $k$ to the BS via passive surface $b$ is given by
\begin{align}
	\boldsymbol{h}_k\left(\boldsymbol{q}_b,\boldsymbol{u}_b\right)=\boldsymbol{V}\left(\boldsymbol{q}_b,\boldsymbol{u}_b\right)\text{diag}\left(\boldsymbol{\theta}_b\right)\boldsymbol{g}_k\left(\boldsymbol{q}_b,\boldsymbol{u}_b\right).
\end{align}
 
As illustrated in Fig. \ref{Fig_main}, by adopting the geometry channel model, the channel $\boldsymbol{g}_k\left(\boldsymbol{q}_b, \boldsymbol{u}_b\right)$ and $\boldsymbol{V}\left(\boldsymbol{q}_b, \boldsymbol{u}_b\right)$ can be expressed as 
 \begin{align}
 	\boldsymbol{g}_{k}\left(\boldsymbol{q}_b, \boldsymbol{u}_b \right)&=\sum_{l=1}^{L_k}a_{k,l}\sqrt{G_{k,l}^{I}\left( \boldsymbol{u}_b\right)}\boldsymbol{t}_{k,l}\left(\boldsymbol{q}_b, \boldsymbol{u}_b \right),\label{eq:eq8}\\
 	\boldsymbol{V}\left(\boldsymbol{q}_b, \boldsymbol{u}_b \right)&=\sum_{p=1}^{P}v_{p}\sqrt{G^{R}_{p}\left( \boldsymbol{u}_b\right)}\boldsymbol{z}_p\boldsymbol{e}^{H}_{p}\left(\boldsymbol{q}_b, \boldsymbol{u}_b \right),\label{eq:eq9}
 \end{align}
where $L_k$ and $P$ denote the number of propagation paths between user $k$ and passive 6DMA surface $b$ and that between surface $b$ and the BS, respectively, $a_{k,l}$ and  $v_{p}$ are the complex-valued gain of the corresponding channels, \(\boldsymbol{z}_p \in \mathbb{C}^{M \times 1}\) represents the steering vector of the BS, and $G_{k,l}^{I}\left( \boldsymbol{u}_b\right)$ $\left(G^{R}_{p}\left( \boldsymbol{u}_b\right)\right)$ denotes the incident (reflective) radiation pattern of each RE in surface $b$ corresponding to the $l$-th path from user $k$ (the $p$-th path to the BS), which is determined by the elevation and azimuth AoAs (AoDs) $\{\phi_{k,l}^{e},\phi_{k,l}^{a}\}$ ($\{\eta_{p}^{e}$, $\eta_{p}^{a}\}$) and the rotation $\boldsymbol{u}_b$ of passive surface $b$. Specifically, we assume that the first paths in (\ref{eq:eq8}) and (\ref{eq:eq9}), corresponding to $l=1$ and $p=1$, are the line-of-sight (LoS) paths; while the remaining $L_k-1$ and $P-1$ paths in (\ref{eq:eq8}) and (\ref{eq:eq9}) are the non-LoS (NLoS) paths.

Let $\boldsymbol{t}_{k,l}\left(\boldsymbol{q}_b, \boldsymbol{u}_b \right)\in\mathbb{C}^{N\times1}$ and $\boldsymbol{e}_{p}\left(\boldsymbol{q}_b, \boldsymbol{u}_b \right)\in\mathbb{C}^{N\times1}$  denote the array response vectors of passive 6DMA surface $b$ from the $l$-th transmit path of user $k$ and that towards the $p$-th reflective path to the BS, respectively, which are obtained as
\begin{align}
	\boldsymbol{t}_{k,l}\left(\boldsymbol{q}_b, \boldsymbol{u}_b \right)&=\left[e^{j\frac{2\pi}{\lambda}\boldsymbol{f}_{k,l}^T\boldsymbol{r}_{b,1}\left(\boldsymbol{q}_b, \boldsymbol{u}_b \right)},\cdots,e^{j\frac{2\pi}{\lambda}\boldsymbol{f}_{k,l}^T\boldsymbol{r}_{b,N}\left(\boldsymbol{q}_b, \boldsymbol{u}_b \right)}\right]^T, \label{eq:eq100}\\
	\boldsymbol{e}_p\left(\boldsymbol{q}_b, \boldsymbol{u}_b\right) &= \left[e^{j\frac{2\pi}{\lambda}\boldsymbol{s}_{p}^T\boldsymbol{r}_{b,1}\left(\boldsymbol{q}_b, \boldsymbol{u}_b \right)},\cdots,e^{j\frac{2\pi}{\lambda}\boldsymbol{s}_{p}^T\boldsymbol{r}_{b,N}\left(\boldsymbol{q}_b, \boldsymbol{u}_b \right)}\right]^T,\label{eq:eq99}
\end{align}
where $\boldsymbol{f}_{k,l}$ and $\boldsymbol{s}_{p}$ are the direction of arrival/departure (DOA/DOD) vectors with unit-norm for the $l$-th path from user $k$ and the $p$-th path to the BS, both w.r.t. the reference position $O$ in Fig. \ref{Fig_main} and $\lambda$ is the carrier wavelength. Specifically,  $\boldsymbol{f}_{k,l}$ is given by 
\begin{align}
	\boldsymbol{f}_{k,l} = \left[\sin(\phi_{k,l}^{e})\cos(\phi_{k,l}^{a}),\sin(\phi_{k,l}^{e})\sin(\phi_{k,l}^{a}),\cos(\phi_{k,l}^{e})\right]^T. \label{eq:eqfkl}
\end{align}
Similarly, $\boldsymbol{s}_p$ can be defined with given $\{\eta_{p}^{e}, \eta_{p}^{a}\}$. Let $\boldsymbol{r}_{b,n}\left(\boldsymbol{q}_b, \boldsymbol{u}_b \right) $ denote the position of the $n$-th RE of 6DMA surface $b$ in the global CCS, $n\in\mathcal{N}$ with $\mathcal{N}\triangleq\left\{1,\cdots,N\right\}$, which can be expressed as 
\begin{align}
	\boldsymbol{r}_{b,n}\left(\boldsymbol{q}_b, \boldsymbol{u}_b \right)=\boldsymbol{q}_b + \boldsymbol{R}\left(\boldsymbol{u}_b\right)\overline{\boldsymbol{r}}_n,\label{eq:eq4}
\end{align}
where $\overline{\boldsymbol{r}}_n\in\mathbb{R}^3$ is the position of the $n$-th RE of 6DMA surface $b$ in its local CCS and $\boldsymbol{R}\left(\boldsymbol{u}_b\right)\in \mathbb{R}^{3\times3}$ is the corresponding rotation matrix generated by $\boldsymbol{u}_b$ as defined in \cite{shao20246d}. Additionally, the array response vector of the $M$-antenna fixed uniform linear array (ULA) at the BS associated with AoA $\psi_p$ from the $p$-th path, denoted by $\boldsymbol{z}_p\in \mathbb{C}^{M\times1}$, can be similarly defined, which is omitted here for brevity.

By denoting $\boldsymbol{q}\triangleq\left[\boldsymbol{q}_1^T,\cdots,\boldsymbol{q}_B^T\right]^T$, $\boldsymbol{u}\triangleq\left[\boldsymbol{u}_1^T,\cdots,\boldsymbol{u}_B^T\right]^T$, and $\boldsymbol{\theta}\triangleq\left[\boldsymbol{\theta}_{1}^{T}, \cdots,\boldsymbol{\theta}_{B}^{T}\right]^T$ as the set of position, rotation, and reflection coefficients of all 6DMA surfaces, the effective channel between user $k$ and the BS via all 6DMA surfaces can be expressed as
\begin{align}
	\boldsymbol{h}_k\left(\boldsymbol{q},\boldsymbol{u}\right)=\boldsymbol{V}\left(\boldsymbol{q},\boldsymbol{u}\right)\text{diag}\left(\boldsymbol{\theta}\right)\boldsymbol{g}_k\left(\boldsymbol{q},\boldsymbol{u}\right),
\end{align}
\noindent \text{with}
\begin{align}	
	\boldsymbol{V}\left(\boldsymbol{q},\boldsymbol{u}\right)&=\left[\boldsymbol{V}\left(\boldsymbol{q}_1,\boldsymbol{u}_1\right),\cdots,\boldsymbol{V}\left(\boldsymbol{q}_B,\boldsymbol{u}_B\right)\right],\\
	\boldsymbol{g}_k\left(\boldsymbol{q},\boldsymbol{u}\right)&=\left[\boldsymbol{g}_k\left(\boldsymbol{q}_1,\boldsymbol{u}_1\right)^T,\cdots,\boldsymbol{g}_k\left(\boldsymbol{q}_B,\boldsymbol{u}_B\right)^T\right]^T.
\end{align}
\subsection{Signal Model}
For the uplink transmission, the BS applies a linear receive beamforming vector $\boldsymbol{w}^H_k\in\mathbb{C}^{M\times 1}$ to decode $x_k$ from user $k$, i.e.,
\begin{align}
	\hat{y}_k=\boldsymbol{w}_k^{H}\boldsymbol{h}_k\left(\boldsymbol{q},\boldsymbol{u}\right)x_k+\boldsymbol{w}_k^{H}\sum_{j=1,j\neq k}^{K}\boldsymbol{h}_j\left(\boldsymbol{q},\boldsymbol{u}\right)x_j+\boldsymbol{w}_k\boldsymbol{c},
\end{align}
where $x_k\sim\mathcal{N}_c(0,P_k)$ is the transmitted data symbol of user $k$, $P_k$ is the corresponding transmit power, and $\boldsymbol{c}\sim\mathcal{CN}\left(0,\sigma^2\textbf{I}\right)$ is the additive white Gaussian noise (AWGN) vector at the BS with $\sigma^2$ being the noise power.
Accordingly, the SINR for decoding the information from user $k$ is given by
\begin{align}
	\gamma_k=\frac{P_k\left|\boldsymbol{w}_k^H\boldsymbol{h}_k\left(\boldsymbol{q},\boldsymbol{u}\right)\right|^2}{\sum_{j=1,j\neq k}^{K}P_j\left|\boldsymbol{w}_k^H\boldsymbol{h}_j\left(\boldsymbol{q},\boldsymbol{u}\right)\right|^2+\sigma^2\left|\boldsymbol{w}_k\right|^2}.\label{sinr}
\end{align}
Accordingly, the corresponding achievable data rate in bits per second per Hertz (bps/Hz) from user $k$ to BS is $\text{log}_2\left(1+\gamma_k\right)$.
\subsection{Problem Formulation}
In this letter, we aim to maximize the achievable sum rate of all users by jointly optimizing the positions, rotations, reflection coefficients of all passive 6DMA surfaces and the receive beamforming matrix at the BS.
Accordingly, the optimization problem is formulated as
\begin{subequations}\label{prob_original2}
	\begin{align}
		(\mathcal{P}_1):~& \mathop{\max}\limits_{\boldsymbol{q}, \boldsymbol{u}, \boldsymbol{W}, \boldsymbol{\theta}} ~ f\left(\boldsymbol{q},\boldsymbol{u}, \boldsymbol{W}, \boldsymbol{\theta} \right)=\sum_{k=1}^{K}\text{log}_2(1+\gamma_k)
		\label{ooP2a}\\
		\mathrm{s.t.} ~ & \boldsymbol{q}_b\in\mathcal{C}, \forall b \in \mathcal{B},\label{ooP2b}\\
		~&\left\|\boldsymbol{q}_b-\boldsymbol{q}_j\right\|^2 \geq d_\text{min}, \forall b,j\in\mathcal{B}, b\neq i,\label{ooP2c}\\
		~&\boldsymbol{n}\left(\boldsymbol{u}_b\right)^T\left(\boldsymbol{q}_j-\boldsymbol{q}_b\right)\leq0, \forall b,j\in\mathcal{B},j\neq b,\label{ooP2z}\\
		~&\boldsymbol{n}\left(\boldsymbol{u}_b\right)^T\boldsymbol{q}_b\geq0, \forall i\in\mathcal{B}\label{ooP2d}\\
		~ & |\theta_{n}|= 1,  n=1,\cdots,NB, \label{ooP2aaaa}
	\end{align}
\end{subequations}
where $\boldsymbol{W}=\left[\boldsymbol{w}_1,\cdots,\boldsymbol{w}_K\right]\in\mathbb{C}^{M\times K}$ denotes the receive beaming matrix. $
	\mathbf{n}(\mathbf{u}_b)=\mathbf{R}(\mathbf{u}_b)\bar{\mathbf{n}}$ with
$\bar{\mathbf{n}}$ being the normal vector of the $b$-th passive 6DMA surface in the local CCS. Constraint (\ref{ooP2b}) guarantees that the movable region of all 6DMA surfaces is located in the given 3D site space $\mathcal{C}$ (e.g., a cube or sphere), (\ref{ooP2c}) constrains the minimum inter-surface distance $d_\text{min}$ to avoid mutual overlap, (\ref{ooP2z}) avoids mutual signal reflection between any two surfaces, (\ref{ooP2d}) prevents 6DMA surfaces from rotating towards the CPU which causes signal blockage \cite{shao20246dma}, and (\ref{ooP2aaaa}) is the unit-modulus constraint for each passive RE.

Note that the formulated problem is challenging to solve due to the coupled variables, non-concave objective function (\ref{ooP2a}), and  non-convex constraints (\ref{ooP2c})-(\ref{ooP2aaaa}).
\section{Proposed Solution}\label{zzz_Part}
In this section, instead of jointly optimizing $\boldsymbol{q}$,  $\boldsymbol{u}$, $\boldsymbol{\theta}$, and $\boldsymbol{W}$ with high computational complexity, an alternating optimization (AO) algorithm is proposed to solve the problem ($\mathcal{P}_1$) more efficiently. As described in Algorithm 1, we first derive the optimal receive beamforming matrix $\boldsymbol{W}$ with antenna position and rotation fixed. Second, the positions $\boldsymbol{q}$ and rotations $\boldsymbol{u}$ of all 6DMA surfaces are optimized via the feasible gradient
descent method (FGD) with given reflection coefficients and beamforming matrix. Finally, the passive beamforming vector $\boldsymbol{\theta}$ of all 6DMA surfaces is optimized via fractional
programming with antenna positions, rotations, and beamforming matrix fixed. 
\subsection{Receive Beamforming Design}
For any given positions $\boldsymbol{q}$, rotations $\boldsymbol{u}$, and reflection coefficients $\boldsymbol{\theta}$ of passive 6DMA surfaces, the receive beamforming at the BS is designed based on the minimum mean square error (MMSE) method, i.e., 
\begin{align}
	\boldsymbol{W}_{MMSE}=\left(\boldsymbol{H}\left(\boldsymbol{q,\boldsymbol{u}}\right)\boldsymbol{P}\boldsymbol{H}\left(\boldsymbol{q,\boldsymbol{u}}\right)^H+\sigma^2\boldsymbol{I}_M\right)^{-1}\boldsymbol{H}\left(\boldsymbol{q,\boldsymbol{u}}\right),
	\label{ooP4a}
\end{align}
where $\boldsymbol{H}\left(\boldsymbol{q,\boldsymbol{u}}\right)\triangleq\left[\boldsymbol{h}_1\left(\boldsymbol{q},\boldsymbol{u}\right),\cdots,\boldsymbol{h}_K\left(\boldsymbol{q},\boldsymbol{u}\right)\right]$ and $\boldsymbol{P}\triangleq\text{diag}\left(P_1,\cdots, P_K\right)$ denote the effective user-BS channel matrix and the diagonal transmit power matrix of $K$ users, respectively.

\subsection{Position and Rotation Optimization}
In each iteration of the proposed AO algorithm, we optimize $\boldsymbol{q}$ and $\boldsymbol{u}$ with given reflection coefficients $\boldsymbol{\theta}$ and beamforming matrix $\boldsymbol{W}$ by updating the position and rotation of one 6DMA surface (i.e., $\boldsymbol{q}_b$ and $\boldsymbol{u}_b, \forall b\in\mathcal{B}$) sequentially with those of the other surfaces fixed.

Given $\boldsymbol{u}_b$, we first utilize the first-order Taylor expansion at the value of $\boldsymbol{q}_b$ in the previous ($t-1$)-th iteration to approximate the non-convex constraint (\ref{ooP2c}) by the following convex constraint:
\begin{align}
	\frac{\left(\boldsymbol{q}_b^{t-1}-\boldsymbol{q}_j\right)^T\left(\boldsymbol{q}_b-\boldsymbol{q}_j\right)}{\left\|\boldsymbol{q}_b^{t-1}-\boldsymbol{q}_j\right\|^2} \geq d_\text{min}.\label{relax_dim}
\end{align}
Thus, the feasible region w.r.t. $\boldsymbol{q}_b$ becomes a convex set. Then, the FGD method is adopted to obtain the solution $\boldsymbol{q}_b$ in the current $t$-th iteration. Note that the FGD method starts with a feasible vector $\boldsymbol{q}_b^{t-1}$ and generates another feasible vector $\boldsymbol{q}_b^t$ as
\begin{align}
	\boldsymbol{q}_b^t = \boldsymbol{q}_b^{t-1} + \kappa^{t-1}\boldsymbol{d}^{t-1}, \label{gradient}
\end{align}
where $\kappa^{t-1}\in\left(0,1\right]$ is the adaptive step size calculated by the Armijio rule and $\boldsymbol{d}^{t-1}$ is a feasible direction. Specifically, $\boldsymbol{d}^{t-1}$ can be chosen as the solution to the following optimization problem.
\begin{subequations}\label{prob_original1}
	\begin{align}
		(\mathcal{P}_2):~& \mathop{\min}\limits_{\boldsymbol{d}^{t-1}} ~ -\nabla f\left(\boldsymbol{q}_b^{t-1}\right)^T\boldsymbol{d}^{t-1}
		\label{ooP7a}\\
		\mathrm{s.t.} ~ & \text{(\ref{ooP2b})}, \text{(\ref{relax_dim})}, \text{(\ref{ooP2z})}, \text{(\ref{ooP2d})},
	\end{align}
\end{subequations}
where $\nabla f\left(\boldsymbol{q}_b^{t-1}\right)$ can be derived by the numerical solution, which is given by
\begin{align}
	\left[\nabla f\left(\boldsymbol{q}_b^{t-1}\right)\right]_j=\lim_{\xi\rightarrow0}\frac{f\left(\boldsymbol{q}_b^{t-1}+\xi\boldsymbol{a}^j\right)-f\left(\boldsymbol{q}_b^{t-1}\right)}{\xi}, 1\leq j\leq 3, \label{nabla}
\end{align}
where $\boldsymbol{a}^j$ is a vector with a one in the $j$-th element and zeros elsewhere. Finally, the non-convex problem ($\mathcal{P}_1$) w.r.t. $\boldsymbol{q}_b$ is transformed into a linear optimization problem ($\mathcal{P}_2$), which can be efficiently solved by linear programming.

Given $\boldsymbol{q}_b$, $\boldsymbol{u}_b$ can be optimized in the similar way. The non-convex constraints (\ref{ooP2z}) and (\ref{ooP2d}) can be relaxed into convex ones by linearly approximating the rotation matrix $\boldsymbol{R}\left(\boldsymbol{u}_b\right)$, which is specified in (52)-(58) in \cite{shao20246d}. Then, the rotation $\boldsymbol{u}_b$ can also be optimized by the FGD method. 

\subsection{Passive Reflection Optimization}
With fixed $\boldsymbol{W}$, $\boldsymbol{q}$ and $\boldsymbol{u}$, we introduce two auxiliary variables $\boldsymbol{\alpha}=\left[\alpha_1,\cdots,\alpha_K\right]$ and $\boldsymbol{\epsilon}=\left[\epsilon_1,\cdots,\epsilon_K\right]$ to convert the objective function of ($\mathcal{P}_1$) into an equivalent form by multiple ratio fractional programming \cite{xue2023multi}, which is written as 
\begin{align}
	\mathop{\min}\limits_{\boldsymbol{\theta}} ~& \boldsymbol{\theta}^{H}\boldsymbol{U}\boldsymbol{\theta}-2\Re\left({\mathbf{v\boldsymbol{\theta}}}\right)
	\label{ooP4z}
\end{align}
\noindent \text{where}
\begin{align}
	\boldsymbol{U}&=\sum_{k=1}^{K}\left|\epsilon_k\right|^2\sum_{j=1}^{K}P_j\boldsymbol{A}_j^H\boldsymbol{w}_k^H\boldsymbol{w}_k\boldsymbol{A}_j,\\
	\mathbf{v}&=\sum_{k=1}^{K}\sqrt{\left(1+\alpha_k\right)P_k}\epsilon_k^*\boldsymbol{w}_k\boldsymbol{A}_k,
\end{align}
with $*$ denoting conjugation operation 
and $\boldsymbol{A}_k=\boldsymbol{V}\left(\boldsymbol{q}, \boldsymbol{u}\right)\text{diag}\left(\boldsymbol{g}_k\left(\boldsymbol{q}, \boldsymbol{u}\right)\right)$. With $\boldsymbol{\theta}$ fixed, the optimal $\alpha_k$ and $\epsilon_k$ are given by $\alpha_k^{opt}=\gamma_k$ and
\begin{align}
	\epsilon_k^{opt}=\frac{\sqrt{\left(1+\alpha_k\right)P_k}\boldsymbol{w}^H_k\boldsymbol{A}_k\boldsymbol{\theta}}{\sum_{j=1}^{K}P_j\left|\boldsymbol{w}^H_k\boldsymbol{A}_j\boldsymbol{\theta}\right|^2+\sigma^2\left|\boldsymbol{w}_k\right|^2}.\label{opt_epp}
\end{align}
Therefore, we can iteratively optimize the $j$-th element of $\boldsymbol{\theta}$, i.e., $\theta_j$ with the other REs' reflection coefficients being fixed. The problem can be formulated as
\begin{subequations}\label{prob_original4}
	\begin{align}
		(\mathcal{P}_3):~&\mathop{\min}\limits_{\theta_{j}} ~ \left|\theta_j\right|^2u_{j,j}-2\Re\left(c_j\theta_j\right)\\
		\mathrm{s.t.} ~ & \left|\theta_j\right| = 1,  j=1,\cdots,NB,
	\end{align}
\end{subequations}
where $u_{j,j}$ is the element in the $j$-th row and $j$-th column of $\mathbf{U}$, and $c_j=v_j-\sum_{i=1,i\neq j}^{NB}\theta^*_{j}u_{j,j}$ with $v_j$ being the $j$-th element of $\mathbf{v}$. Then, the optimal $\theta_j$ is easily obtained as $\theta_j^{opt}=e^{j\beta^{opt}_j}$ with $\beta^{opt}_j=\angle {c_j^*}$. The other reflection coefficients $\theta_i, i\neq j$ can be updated in the similar way.
The proposed AO for solving problem ($\mathcal{P}_1$) is summarized in Algorithm 1.
\begin{algorithm}[!t]
	\caption{Proposed AO Algorithm for Solving ($\mathcal{P}_1$)}
	\label{alg:AOA}
	\renewcommand{\algorithmicrequire}{\textbf{Input:}}
	\renewcommand{\algorithmicensure}{\textbf{Output:}}
	\begin{algorithmic}[1]	
		\STATE  Input $\left\{\boldsymbol{q}_b^0\right\}_{b=1}^B$, $\left\{\boldsymbol{u}_b^0\right\}_{b=1}^B$, $\left\{\boldsymbol{\theta}_b\right\}_{b=1}^B$, and the iteration numbers $T_1$ and $T_2$.
		\FOR{$t=1:1:T_1$}
		\STATE Update $\boldsymbol{W}^{\left(t\right)}$ via (\ref{ooP4a}).
	    \FOR{$i=1:1:T_2$}
	    \FOR{$b=1:1:B$}
		\STATE  Calculate $\nabla f\left(\boldsymbol{q}_b^{i-1}\right)$ via (\ref{nabla});
		\STATE  Obtain $\boldsymbol{d}^{i-1}$ by solving problem ($\mathcal{P}_3$);
		\STATE  Update $\left(\boldsymbol{q}_b^{i}\right)$ via ($\ref{gradient}$);
 	    \ENDFOR
 	    \FOR{$b=1:1:B$}
 	    \STATE  Calculate $\nabla f\left(\boldsymbol{u}_b^{i-1}\right)$ via (62) in \cite{shao20246d};
 	    \STATE  Obtain $\boldsymbol{\tilde{\boldsymbol{u}}}^{i-1}$ by solving problem (\text{P3-b-2}) according to \cite{shao20246d};
 	    \STATE  Update $\left(\boldsymbol{u}_b^{i}\right)$ via (60) in \cite{shao20246d};
 	    \ENDFOR
 	    \FOR{$j=1:1:NB$}
 	    \STATE  Update $\theta_j$ by solving Problem ($\mathcal{P}_2$);
 	    \ENDFOR
		\ENDFOR
		\ENDFOR
	    \STATE	Update $\boldsymbol{W}$ via (\ref{ooP4a}).
		\RETURN $\boldsymbol{W}$, $\boldsymbol{\theta}$, $\boldsymbol{q}$, and $\boldsymbol{u}$.
	\end{algorithmic}
\end{algorithm}
\section{Numerical Results}
To evaluate the performance of the proposed scheme, we consider the following simulation setup. The carrier wavelength is $\lambda=0.125$ m. The number of REs for each 6DMA surface is set to be equal. The number of BS antennas is $M=6$ and the number of users is $K=6$. Moreover, $d_{\text{min}}$ is set as $\frac{\sqrt{2}}{2}\lambda+\frac{\lambda}{10}$. For the channel model, we set $L_k=2$, with $a_{k,1}\sim \mathcal{CN}\left(0, 4\times10^{-6}\right)$ and $a_{k,l}\sim\mathcal{CN}\left(0, 1\times10^{-6}\right), \forall l\in \mathcal{L}_k \setminus \{1\}$. In addition, $\phi_{k,l}^{e}$ and $\phi_{k,l}^{a}$ are randomly and uniformly generated in the intervals $\left[0,\pi/2\right]$ and $\left[\pi, 3\pi/2\right]$, respectively. $P$ is set as 6 and $v_{p}$ has the same distribution as $a_{k,l}$. $\psi_p$, $\eta_{p}^{e}$, and $\eta_{p}^{a}$ are randomly generated in intervals $\left[0,\pi/2\right]$, $\left[\pi/2,\pi\right]$ and $\left[-\pi/2,0\right]$, respectively. The transmit power of each user $\left\{P_k\right\}_{k=1}^K$ is set to be equal. The noise power $\sigma^2$ is set to -80 dBm. To investigate the effectiveness of the proposed passive 6DMA-assisted multiuser uplink communication, we consider the following schemes (with the same total number of REs) for performance comparison: 1) Distributed passive 6DMA with multiple surfaces, i.e., $B=4, N=4$; 2) Centralized passive 6DMA with a single surface, i.e., $B=1$, $N=16$; 3) Traditional fixed IRS with $N=16$.

According to \cite{Zheng}, the directive radiation pattern of each RE for incident signals  is given by 
\begin{align}
	G_{k,l}^{I}\left(\boldsymbol{u}_b\right) = \begin{cases}
		\frac{A\times\langle \boldsymbol{n}\left(\boldsymbol{u}_b\right), -\boldsymbol{f}_{k,l} \rangle}{\lambda^2/4\pi},& \text{if} \ \langle \boldsymbol{n}\left(\boldsymbol{u}_b\right), -\boldsymbol{f}_{k,l} \rangle > 0,\\
		0, & \text{otherwise},
	\end{cases}\label{eq:eq6}
\end{align}
where $A$ denotes the area of each RE, usually set as $\left(\lambda/2\right)^2$, $\overline{\boldsymbol{n}}\triangleq\left[1,0,0\right]^T$ is the outward normal vector of 6DMA surface $b$ in its local CCS, and $\langle \boldsymbol{n}\left(\boldsymbol{u}_b\right), -\boldsymbol{f}_{k,l} \rangle$ computes the cosine of angle between the DOA vector of incident signal $\boldsymbol{f}_{k,l}$ and outward normal vector of 6DMA surface $b$ in the global CCS. 
Similarly, the directive reflective radiation pattern \( G_{p}^{R}\left(\boldsymbol{u}_b\right) \) can be obtained by substituting \( \boldsymbol{f}_{k,l} \) in \eqref{eq:eq6} with \( \boldsymbol{s}_{p} \). Furthermore, we consider the case of a half-space isotropic radiation pattern for both $G_{k,l}^{I}\left(\boldsymbol{u}_b\right)$ and $G_{p}^{R}$ as a comparison, which is obtained by replacing the term \(\langle \cdot \rangle\) in the numerator of the directive radiation pattern with the constant 2 \cite{Zheng}.

\begin{figure}[!t]
	\centering
	\includegraphics[width=0.35\textwidth]{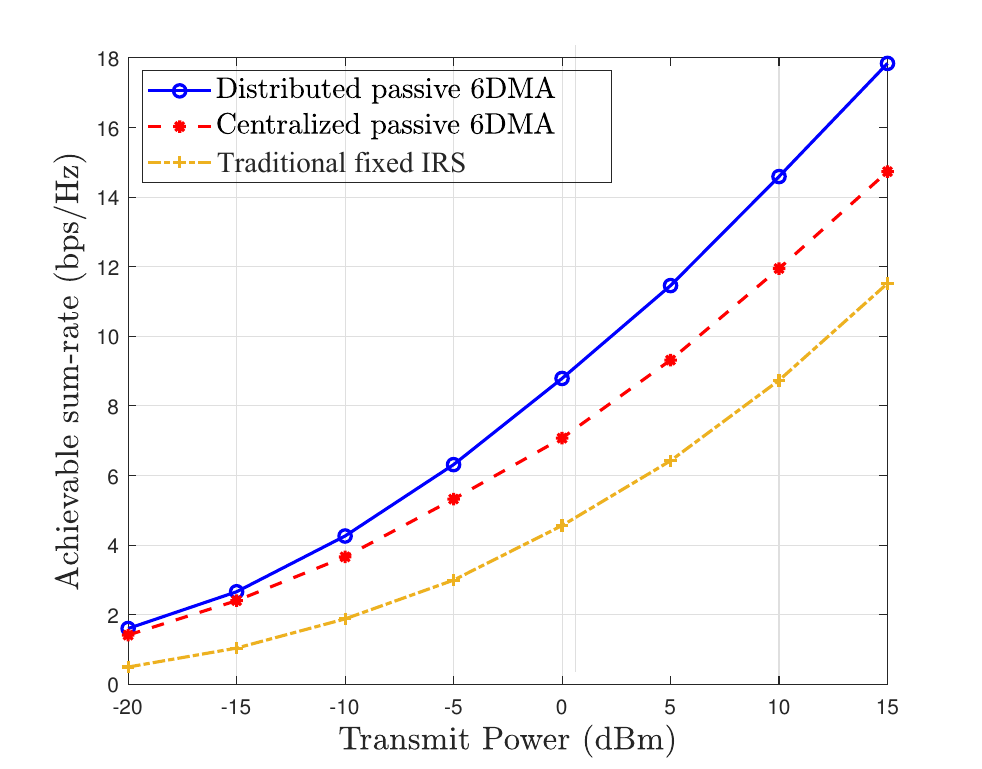}
	\caption{Sum rate versus transmit power in the case of directive radiation pattern of REs.}
	\label{Fig2_directive}
		\vspace{-0.42cm}
\end{figure}

In Fig. \ref{Fig2_directive}, we show the sum rate versus user transmit power in the case of directive radiation pattern of REs for different schemes. It is observed that the proposed passive 6DMA schemes (either distributed or centralized) outperform the fixed IRS scheme, where the performance gaps become  more substantial as transmit power increases. This is due to the extra DoFs introduced by the passive 6DMA with position and rotation adjustment. Although the total number of REs is the same, the passive 6DMA with distributed surfaces performs better than the benchmark scheme with a centralized surface. This is due to the increased flexibility in position and rotation adjustment with distributed surfaces, which also helps reduce inter-user interference. However, this practically requires more hardware cost and higher complexity for controlling the movement of individual 6DMA surfaces.
\begin{figure}[!t]
	\centering
	\includegraphics[width=0.35\textwidth]{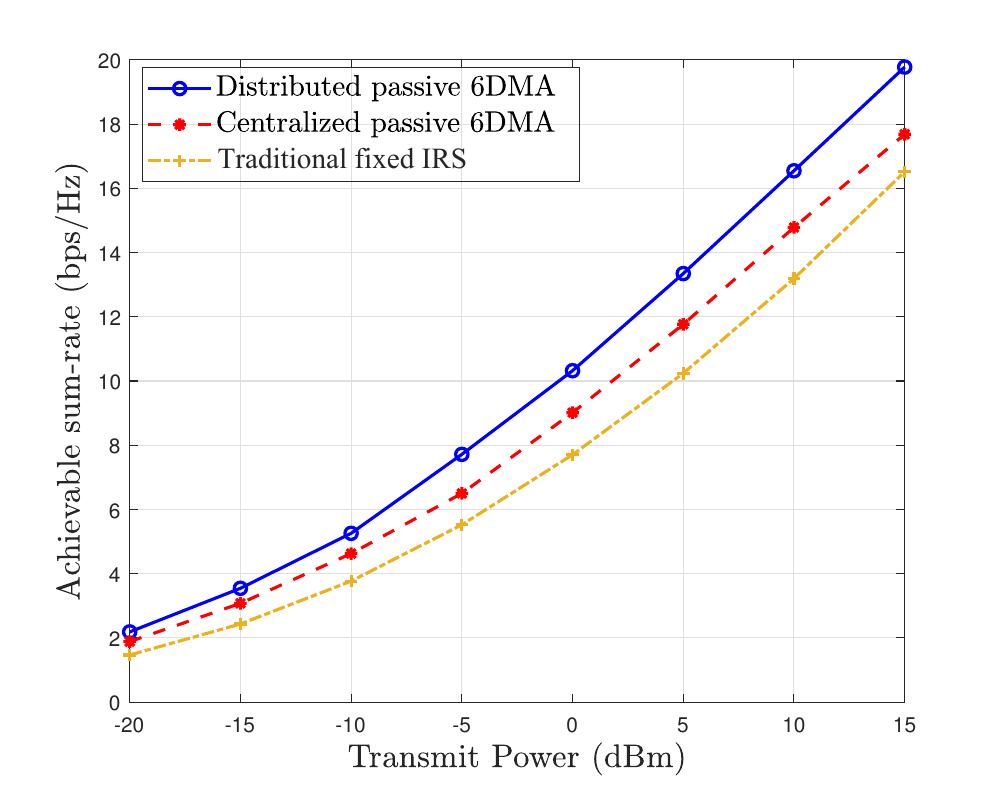}
	\caption{Sum rate versus transmit power in the case of half-space isotropic radiation pattern of REs.}
	\label{Fig3_isotropic}
		\vspace{-0.39cm}
\end{figure}

In Fig. \ref{Fig3_isotropic}, we show the sum rate versus user transmit power in the case of  half-space isotropic radiation pattern of REs for different schemes. It is observed that the performance gaps become smaller among different schemes as compared to the case with directive radiation pattern of REs. For example, the sum rate with the fixed-IRS scheme with half-space isotropic radiation pattern increases about 5 bps/Hz as compared with that of directive radiation pattern at the user transmit power of 15 dBm. In contrast, the sum rate with the proposed distributed/centralized passive 6DMA schemes increases only 2 bps/Hz and 3 bps/Hz, respectively. The above results indicate that the (practical) directive radiation pattern of REs has less rate loss as compared to the (ideal) half-space isotropic radiation pattern when the distributed passive 6DMA is employed, as the surface rotation can make better use of the directive radiation pattern of REs for desired signal enhancement and inter-user interference suppression.

\section{Conclusion}
In this letter, we studied a new passive 6DMA-assisted multiuser uplink communication system by exploiting the new design  DoFs through 6DMA surfaces' position and rotation adjustment. We formulated an optimization problem aiming to maximize the achievable sum rate of users by jointly optimizing the surfaces' position, rotation, and reflection coefficients under practical constraints. Numerical results were shown to evaluate the performance of the proposed passive 6DMA system in different configurations (distributed versus centralized) and validated their advantages over the traditional fixed IRS for both directive and half-space isotropic radiation patterns of REs.

\bibliographystyle{ieeetr}
\bibliography{reference}
\end{document}